\documentclass[a4paper,twocolumn,english,nosuperscriptaddress,showpacs,prl]{revtex4}

\usepackage[T1]{fontenc}
\usepackage[latin9]{inputenc}
\usepackage{babel}
\usepackage{amsmath}
\usepackage{amssymb}
\usepackage{times}
\usepackage{graphicx}

\usepackage[unicode=true,pdfusetitle,
 bookmarks=false,bookmarksnumbered=false,bookmarksopen=false,
 breaklinks=false,pdfborder={0 0 1},backref=line,colorlinks=false]
 {hyperref}

\begin{document}

\title{A proposal to implement a quantum delayed choice experiment assisted by a cavity QED}

\author{N. G. de Almeida}
\email{norton@if.ufg.br}
\affiliation{Instituto de Física, Universidade Federal de Goiás, 74.001-970, Goiânia,
Goiás, Brazil}
\author{A T. Avelar}
\email{ardiley@if.ufg.br}
\affiliation{Instituto de Física, Universidade Federal de Goiás, 74.001-970, Goiânia,
Goiás, Brazil}
\author{W. B. Cardoso}
\email{wesleybcardoso@gmail.com.br}
\affiliation{Instituto de Física, Universidade Federal de Goiás, 74.001-970, Goiânia,
Goiás, Brazil}

\begin{abstract}
We propose a scheme feasible with current technology to implement
a quantum delayed-choice experiment in the realm of cavity QED. Our
scheme uses two-level atoms interacting on and off resonantly with
a single mode of a high Q cavity. At the end of the protocol, the
state of the cavity returns to its ground state, allowing new sequential
operations. The particle and wave behavior, which are verified in
a single experimental setup, are postselected after the atomic states
are selectively detected. 
\end{abstract}


\maketitle

\emph{Introduction}. Recently, the quantum version of Wheeler's delayed-choice
experiment (QDCE) was proposed \cite{Terno} and experimentally demonstrated
for photons \cite{Tang2012,Popescu,Kaiser} as well as for spins \cite{Auccaise,Mahesh}.
Different from the classical version \cite{Wheeler}, which also has
been tested in a number of papers \cite{DSE}, in the QDCE the detecting
device can also occupy a quantum state. In general, the goal of delayed-choice
experiment is to test the complementarity principle, which states
that the wave-like (WL) behavior revealed by the appearance of interference
patterns and particle-like (PL) behavior are complementary and mutually
exclusive, thus needing two distinct experimental arrangements to
be verified. However, the quantum version of QDCE enables one to measure
complementary phenomena with a single experimental setup by postselecting
the WL or PL behavior, thus pointing to a redefinition of the complementarity
principle, such that, instead of complementarity of experimental setups
according to Bohr's view, we have complementarity of experimental
data \cite{Terno}.

Another interesting feature of the QDCE is to prove that there are
no consistent local hidden-variable (LHV) theories having ``particle\textquotedblright{}\ and
``wave\textquotedblright{}\ as realistic properties. To this prove,
the operational definition for ``wave\textquotedblright{}\ or ``particle\textquotedblright{}\ was
given as the ``ability\textquotedblright{}\ or ``inability\textquotedblright{}\
to produce interference \cite{Terno}. This operational definition
was considered further by Filgueiras \textit{et al. }\cite{Filgueiras}
to show incompatibility between quantum and LHV theories even when
arbitrary amounts of white noise is included into the optical QDCE.
In this paper, we propose a simplified scheme to realize the analog
of the QDCE also in the domain of cavity QED. Our scheme uses only
two-level atoms interacting on and off resonantly with a single mode
of a cavity, which is disregarded after the interaction, and selective
atomic state detectors. The whole setup we are proposing are well
known from experiments on cavity QED \cite{haroche}, thus being completely
feasible using current technology.

\begin{figure}[tb]
\includegraphics[width=0.9\columnwidth]{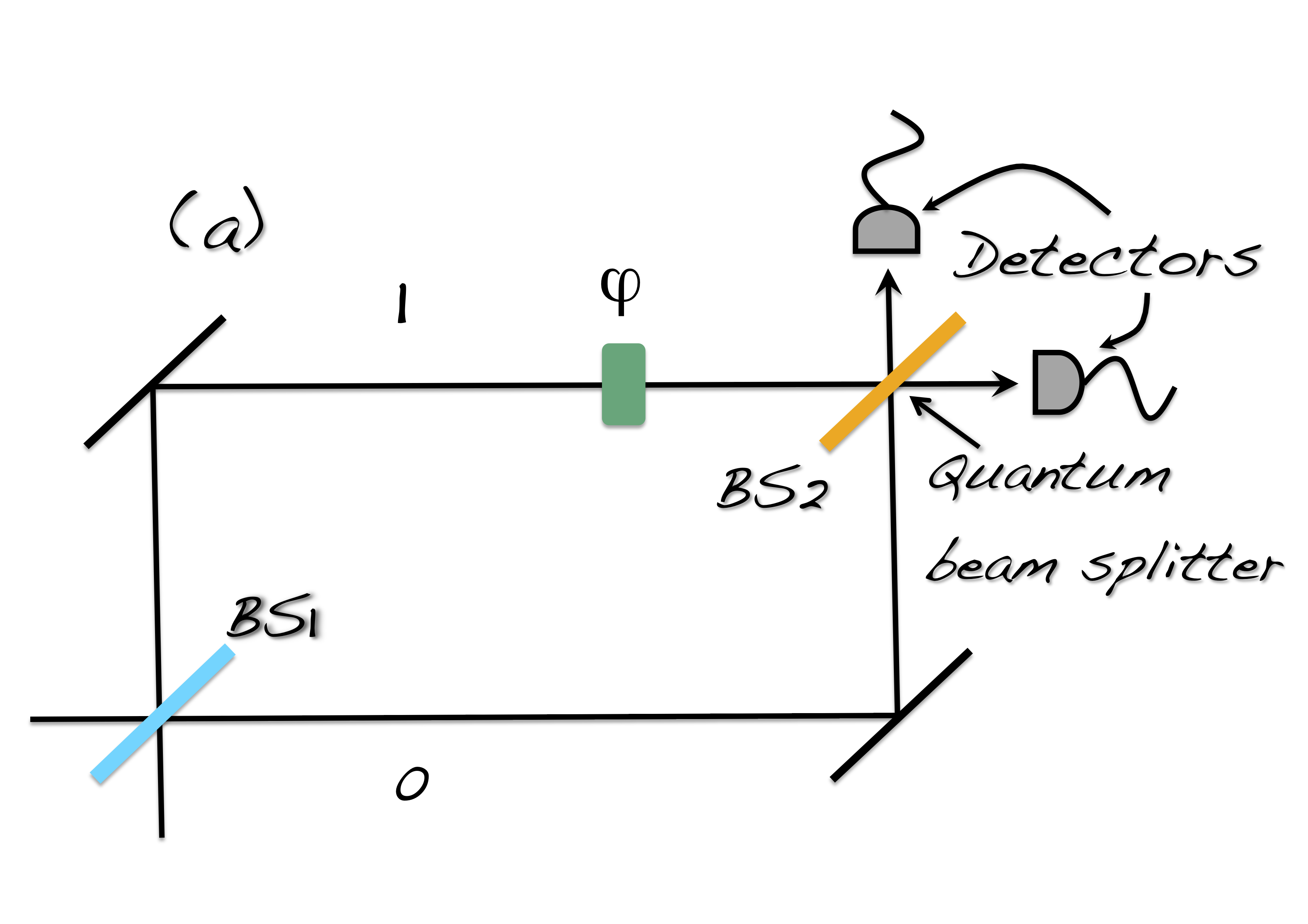} \includegraphics[width=0.5\columnwidth]{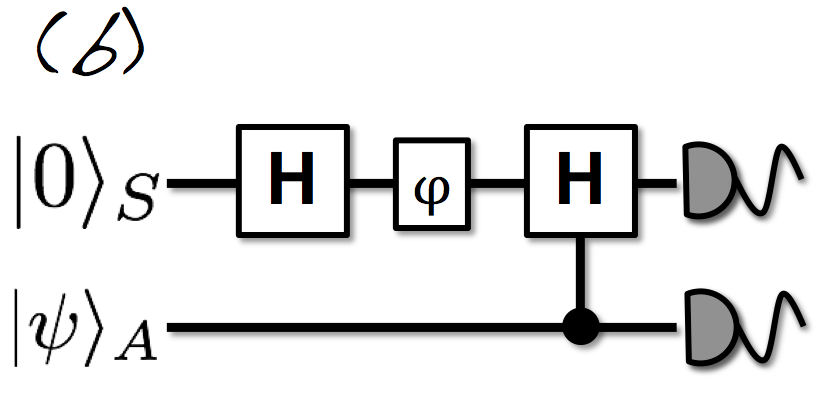}
\caption{(Color online) (a) Schematic diagram of the Mach-Zehnder interferometer
with a \textit{quantum} beam splitter $BS_{2}$. (b) The quantum circuit
that describes the evolution of the ancilla and the photon in the
interferometer \cite{Terno}. The ancilla is the qubit in the lower
line of the circuit, while the qubit inside the interferometer is
in the upper line. The state of the ancilla is given by $|\psi\rangle_{A}=\cos\alpha|0\rangle_{A}+\sin\alpha|1\rangle_{A}$.
$H$ is the Hadamard gate and $\varphi$ is a gate that creates the
phase difference $\varphi$ between the paths $0$ and $1$. The interferometer
is closed for $\alpha=\pi/2$ and open for $\alpha=0$. For any other
value, $0<\alpha<\pi/2$, the interferometer is in a coherent superposition
of being closed and open.}
\label{Fig1} 
\end{figure}

In the Mach-Zehnder interferometer, as shown in Fig. \ref{Fig1}(a),
interference patterns giving rise to WL behavior appear in detectors
placed on paths $1$ and $0$ when the interferometer is closed, i.e.,
when the second beam splitter $\mathcal{BS}_{2}$ is present. Otherwise,
if the second beam splitter is absent the experiment reveals which-path
information, and a PL behavior is observed. In the language of the
complementarity principle, if we want to observe the wave aspect of
the photon, we must consider the closed interferometer (with $\mathcal{BS}_{2}$
present), whereas to observe the particle nature of the photon we
must consider the open interferometer (removing $\mathcal{BS}_{2}$).
Note, therefore, that these two different experimental arrangements
are complementary in the sense that each choice determines beforehand
the statistics of the results by the experimenter's decision. This
is a classical experiment, in the sense that the interferometer has
only two states, open or closed.

In the quantum extension of the delayed choice experiment \cite{Terno},
the second beam splitter $\mathcal{BS}_{2}$ in Fig. \ref{Fig1}(a)
is in a coherent superposition of being present and absent and is
now controlled by a quantum device, referred to as the ancilla system,
which allows the beam splitter to be in a superposition of being present
and absent. Fig. \ref{Fig1}(b) shows the quantum circuit to describe
the evolution of the system through the interferometer. Considering
as the initial state of the system- ancilla 
\begin{equation}
|\psi\rangle_{\mathcal{SA}}^{in}=|0\rangle_{\mathcal{S}}\otimes\left[\cos\alpha|0\rangle_{\mathcal{A}}+\sin\alpha|1\rangle_{\mathcal{A}}\right]\text{,}\label{initial_state}
\end{equation}
then, after the action of the second beam splitter the final system-ancilla
state is 
\begin{equation}
|\psi\rangle_{\mathcal{SA}}^{out}=\cos\alpha|p\rangle_{\mathcal{S}}|0\rangle_{\mathcal{A}}+\sin\alpha|w\rangle_{\mathcal{S}}|1\rangle_{\mathcal{A}}\text{,}\label{pure}
\end{equation}
where $|p\rangle_{\mathcal{S}}=\left(|0\rangle_{\mathcal{S}}+e^{i\varphi}|1\rangle_{\mathcal{S}}\right)/\sqrt{2}$
accounts for PL statistical behavior of the photon, while $|w\rangle_{\mathcal{S}}=e^{i\varphi/2}\left(\cos\left(\varphi/2\right)|0\rangle_{\mathcal{S}}-i\sin\left(\varphi/2\right)|1\rangle_{\mathcal{S}}\right)$
accounts for its WL behavior, and states $|1\rangle_{\mathcal{S}}$
and $|0\rangle_{\mathcal{S}}$ label the interferometric paths $1$
and $0$, respectively. Note that the transformation employed by the
second beam splitter is coherently controlled by the ancillary system,
i.e., the ancilla in the state $|0\rangle_{\mathcal{A}}$ corresponds
to the absence of the second beam splitter, modeling an open interferometer.
On the other hand, the ancilla in the state $|1\rangle_{\mathcal{A}}$
corresponds to $\mathcal{BS}_{2}$ present, then modeling a closed
interferometer. Since $\mathcal{BS}_{2}$ is now a quantum system,
its state is not limited to be present or absent, but can be in any
superposition of $|0\rangle_{\mathcal{A}}$ and $|1\rangle_{\mathcal{A}}$,
meaning that the interferometer can be cast in an arbitrary superposition
of being open and closed \cite{Terno}. An interesting behavior displaying
a continuous morphing between PL and WL behavior is verified by varying
the parameter $\alpha$.

If we now use the computational basis $\left(00,01,10,11\right)$
in the ${\mathcal{S}\otimes\mathcal{A}}$ space, it is straightforward
to calculate this final joint probability distribution 
\begin{eqnarray}
P\left(S,A\right) & = & \left[\frac{1}{2}\cos^{2}\alpha,\sin^{2}\alpha\cos^{2}\frac{\varphi}{2},\right.\notag\\
 &  & \left.\frac{1}{2}\cos^{2}\alpha,\sin^{2}\alpha\sin^{2}\frac{\varphi}{2}\right]\text{,}\label{statpuro}
\end{eqnarray}
with $S$ $\in$ $\mathcal{S}$ and $A$ $\in$ $\mathcal{A}$ representing
the measurement outcomes in the computational basis. As demonstrated
in Ref.\cite{Terno,Filgueiras}, there is no LHV theory that reproduces
this set of probabilities, even in the presence of an arbitrary amount
of noise.

\textit{Controlled interactions. }To perform the QDCE assisted by
a high$-Q$ cavities we will need the following operators:{\large \ }
\begin{equation}
H_{on}=\hbar g\left(\sigma^{-}a^{\dagger}+\sigma^{+}a\right)\text{,}\label{hon}
\end{equation}

\begin{equation}
H_{off}=\frac{\hbar g^{2}}{\delta}a^{\dagger}a\sigma_{ee}\text{,}\label{hoff}
\end{equation}

\begin{equation}
R=\hbar(\lambda\sigma^{+}+\lambda^{\ast}\sigma^{-})\text{.}\label{zr}
\end{equation}
Eq.(\ref{hon}) is the usual Jaynes-Cummings model \cite{scully}
and describes a resonant atom-field interaction. Here $a^{\dagger}$
and $a$ stands for creation and annihilation operators, respectively,
the two-level atom is described by the lowering ($\sigma^{-}$) and
raising ($\sigma^{+}$) Pauli operators, and $g$ is the atom-field
coupling parameter. Eq.(\ref{hoff}) stands for the dispersive atom-field
interaction \cite{scully} and can be implemented via Stark shift;
$\delta=\left(\omega-\omega_{0}\right)$ is the detuning between the
field frequency $\omega$\ and the atomic{\large \ } frequency $\omega_{0}$,
and $\sigma_{ee}=\left\vert e\right\rangle \left\langle e\right\vert $.
Eq.(\ref{zr}) represents{\large \ }the Ramsey zone \cite{haroche},
where $\lambda=\left\vert \lambda\right\vert \exp(i\chi)$ is the
coupling parameter, which can be adjusted to produce arbitrary rotations
in the internal atomic states $\left\vert g\right\rangle $ and $\left\vert e\right\rangle $
of the two-level atom. Using the operators defined in Eqs. (\ref{hon})-(\ref{zr}),
\ it is straightforward to verify the following evolutions

\begin{equation}
R\left\{ \begin{array}{c}
\left\vert e\right\rangle \rightarrow\cos\theta\left\vert e\right\rangle -i\exp(i\chi)\sin\theta\left\vert g\right\rangle \\
\left\vert g\right\rangle \rightarrow\cos\theta\left\vert g\right\rangle +i\exp(i\chi)\sin\theta\left\vert e\right\rangle 
\end{array}\right.\text{,}\label{zrevol}
\end{equation}
where $\theta=\left\vert \lambda\right\vert gt$, $t$ being the interaction
time,

\begin{equation}
U_{on}\left\{ \begin{array}{c}
\left\vert g\right\rangle \left\vert 0\right\rangle \rightarrow+\left\vert g\right\rangle \left\vert 0\right\rangle \\
\left\vert e\right\rangle \left\vert 0\right\rangle \rightarrow-i\left\vert g\right\rangle \left\vert 1\right\rangle \\
\left\vert g\right\rangle \left\vert 1\right\rangle \rightarrow-i\left\vert e\right\rangle \left\vert 0\right\rangle 
\end{array}\right.\text{,}\label{uonevol}
\end{equation}

\begin{equation}
U_{off}\left\{ \begin{array}{c}
\left\vert e\right\rangle \left\vert 1\right\rangle \rightarrow\exp(i\vartheta)\left\vert e\right\rangle \left\vert 1\right\rangle \\
\left\vert g\right\rangle \left\vert 1\right\rangle \rightarrow\left\vert g\right\rangle \left\vert 1\right\rangle \\
\left\vert e\right\rangle \left\vert 0\right\rangle \rightarrow\left\vert e\right\rangle \left\vert 0\right\rangle \\
\left\vert e\right\rangle \left\vert 0\right\rangle \rightarrow\left\vert e\right\rangle \left\vert 0\right\rangle 
\end{array}\right.\text{,}\label{uoffevol}
\end{equation}
where the above evolutions $U_{on}$ and $U_{off}$ \ are obtained
by adjusting the interaction times as $gt=\pi/2$ and $g^{2}t/\delta=\vartheta$\ from
$U_{on}=\exp\left[-\frac{i}{\hbar}H_{on}t\right]$ and $U_{off}=\exp\left[-\frac{i}{\hbar}H_{off}t\right]$,
respectively. The Hadamard gate $H$ indicated in the circuit of Fig.
\ref{Fig1}(b) is achieved by properly adjusting $\theta$ and $\chi$,
as we shall see in the next Section.

\begin{figure}
\includegraphics[width=0.9\columnwidth]{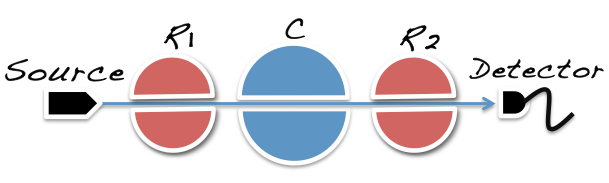} 
\caption{(Color online) Experimental setup to implement the quantum delayed
choice. It consists of a Source of two-level atoms, Ramsey zones (R1
and R2), one microwave cavity C, and selective atomic state detectors.}
\label{Fig2} 
\end{figure}

\begin{figure}
\includegraphics[width=1\columnwidth]{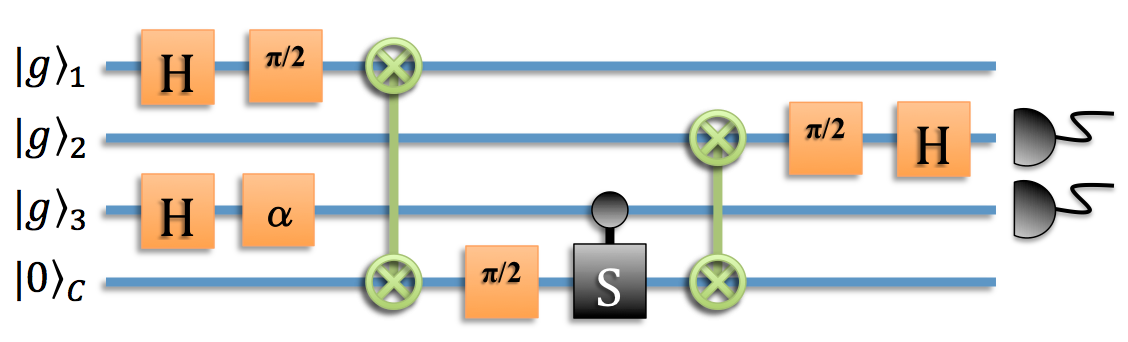}
\caption{(Color online) Effective circuit corresponding to our proposal to
implement QDCE in cavity QED. In this circuit are indicated the Hadamard
gate, the Swap gate, and the controlled phase gate. $\pi/2$ indicates
the relative phase in atomic states. At the final stage, both the
cavity $C$ and the atom $A_{1}$ end in their ground state. While
the atom $A_{1}$ is disregarded, the cavity $C$ is now prompt for
new sequential operations. The Hadamard gate combined with a $\pi/2$
gate can be achieved through a single controlled operation. The same
is true for the SWAP and $\pi/2$ gates.}
\label{Fig4} 
\end{figure}

\textit{Experimental setup. }Our proposal to implement QDCE in the
realm of cavity QED is sketched in Fig. \ref{Fig2}. To reproduce
the state of Eq.(\ref{pure}) and its probability distribution, Eq.(\ref{statpuro}),
using the controlled operations as described in the previous section,
consider the effective circuit as shown in Fig. \ref{Fig4}.

Initially, the system, as indicated in the circuit of Fig. \ref{Fig4},
is in the ground state. In the first step, Rydberg atoms $A_{1}$
and $A_{3}$ (here to be the ancilla) emitted by the source in their
ground state, cross the Ramsey zones $R_{1}$ and $R_{2}$ and are
prepared in the states $(|g\rangle_{1}+i|e\rangle_{1})/\sqrt{2}$
and $\cos\alpha|g\rangle_{A}+\sin\alpha|e\rangle_{A}$, respectively,
according to Eq.(\ref{zrevol}). Note that this operation, indicated
by a Hadamard gate followed by a $\pi/2$ gate in Fig. \ref{Fig4},
is achieved through a single controlled operation Eq(\ref{hon}),
such that before the first SWAP gate the whole state is:
\begin{eqnarray*}
|g,g,,g,0\rangle_{123C} & \rightarrow & \frac{(|g\rangle_{1}+i|e\rangle_{1})}{\sqrt{2}}|g\rangle_{2}\\
 & \otimes & \left(\cos\alpha|g\rangle_{3}+\sin\alpha|e\rangle_{3}\right)|0\rangle_{C}\text{,}
\end{eqnarray*}
where the atom $A_{2}$, indicated in its ground state $|g\rangle_{2},$
is to be the system encoding the WL and PL statistics. The SWAP gate,
followed by a $\pi/2$ gate, is accomplished at once by the Hamiltonian
Eq.(\ref{uonevol}) when $A_{1}$ traverses the cavity $C$ and interacts
resonantly with the field mode in the vacuum state $|0\rangle_{2}$
with the time interaction adjusted to $gt=\pi/2$:
\[
|g\rangle_{1}|g\rangle_{2}\left(\cos\alpha|g\rangle_{3}+\sin\alpha|e\rangle_{3}\right)\frac{(|0\rangle_{C}+|1\rangle_{C})}{\sqrt{2}}\text{.}
\]
The next step is the controlled phase gate, which is accomplished
when the atom $A_{3}$ interacts off resonantly with the cavity $C$\ 
according to Hamiltonian Eq.(\ref{uoffevol}) with an arbitrary parameter
$g^{2}t/\delta=\vartheta$:
\begin{align*}
\frac{1}{\sqrt{2}}|g\rangle_{1}|g\rangle_{2}[\cos\alpha|g\rangle_{3}|0\rangle_{C}+\sin\alpha|e\rangle_{3}|0\rangle_{C}\\
+\cos\alpha|g\rangle_{3}|1\rangle_{C}+\sin\alpha\exp(i\vartheta)|e\rangle_{3}|1\rangle_{C}]\text{.}
\end{align*}
Now, the second SWAP gate between the atom $2$ and cavity $C$ (followed
by the $\pi/2$ gate on the path of atom $2$) is achieved in the
same way as done previously: atom 2 interacts resonantly with the
cavity field, Eq.(\ref{uonevol}), with $gt=\pi/2$:
\begin{align*}
\frac{1}{\sqrt{2}}|g\rangle_{1}[\cos\alpha|g\rangle_{3}|g\rangle_{2}+\sin\alpha|e\rangle_{3}|g\rangle_{2}\\
-i\cos\alpha|g\rangle_{3}|e\rangle_{2}-i\sin\alpha\exp(i\vartheta)|e\rangle_{3}|e\rangle_{2}]|0\rangle_{C}\text{.}
\end{align*}
in the last step of the circuit, a Hadamard gate is achieved when
atom $2$ goes through a Ramsey zone with $\theta=\frac{\pi}{4}$
and $\chi=\pi/2$. Note that the cavity state returns to its initial
ground state, allowing a new sequential operation. Therefore, before
detection and disregarding both the cavity $C$ and atom $A_{1},$
the final state is
\[
|\psi\rangle_{23}=\cos\alpha|p\rangle_{2}|g\rangle_{3}+\sin\alpha|w\rangle_{2}|e\rangle_{3}\text{,}
\]
where 
\begin{eqnarray}
|p\rangle_{2} & = & e^{i\frac{\pi}{4}}\frac{(|g\rangle_{2}+i|e\rangle_{2})}{\sqrt{2}}\notag\\
|w\rangle_{2} & = & e^{i\frac{\phi}{2}}\left[\cos\frac{\phi}{2}|g\rangle_{2}-i\sin\frac{\phi}{2}|e\rangle_{2}\right]\notag
\end{eqnarray}
and $\phi=\frac{\vartheta+\pi}{2}$. The morphing behavior between
wave and particle is thus verified by varying the continuous parameter
$\alpha$.

\textit{Conclusion. }We have proposed a feasible experiment to implement
a quantum delayed choice experiment (QDCE) in the realm of cavity
QED. Our proposal relies on controlled unitary operations, such as
the routinely implemented in cavity QED experiments using two-level
atoms interacting on and off resonantly with a single mode of a cavity
field, plus selective atomic state detectors. Given the technology
currently achieved for high Q cavities, in which a photon lifetime
reaches $0.1s$ \cite{Gleyzes07}, we have disregarded losses both
in atomic and cavity field states.

\begin{acknowledgments}
The authors thank CNPq (National Counsel of Technological and Scientific
Development) and INCT-IQ (National Institute of Science and Technology
of Quantum Information), Brazilian Agencies, for financial support. 
\end{acknowledgments}


\begin{thebibliography}{99}
\bibitem{Terno} R. Ionicioiu and D. R. Terno, Phys. Rev. Lett. \textbf{107},
230406 (2011).

\bibitem{Tang2012} J.-S. Tang, Y.-L. Li, X.-Y. Xu, G.-Y. Xiang, C.-F.
Li, and G.-C. Guo, Nature Photon. \textbf{6}, 600 (2012).

\bibitem{Popescu} A. Peruzzo, P. Shadbolt, N. Brunner, S. Popescu,
and J. L. O'Brien, Science \textbf{338}, 634 (2012).

\bibitem{Kaiser} F. Kaiser, T. Coudreau, P. Milman, D. B. Ostrowsky,
and S. Tanzilli, Science \textbf{338}, 637 (2012).

\bibitem{Auccaise} R. Auccaise, R. M. Serra, J. G. Filgueiras, R.
S. Sarthour, I. S. Oliveira, and L. C. C\'eleri, Phys. Rev. A \textbf{85},
032121 (2012).

\bibitem{Mahesh} S. S. Roy, A. Shukla, and T. S. Mahesh, Phys. Rev.
A \textbf{85}, 022109 (2012).

\bibitem{Wheeler} J. A. Wheeler, in \emph{Mathematical Foundations
of Quantum Mechanics}, A. R. Marlow, ed. (Academic, New York, 1978).

\bibitem{DSE} S. S. Roy, A. Shukla, and T. S. Mahesh, Phys. Rev.
A \textbf{85}, 022109 (2012).

\bibitem{Filgueiras} J. G. Filgueiras, R. S. Sarthour, A. M. S. Souza,
I. S. Oliveira, R. M. Serra, and L. C. Céleri, arXiv:1208.0802 (2012).

\bibitem{scully}M. O. Scully and M. S. Zubary, \emph{Quantum Optics},
Cambridge Univ. press, (1997).

\bibitem{haroche} J. M. Raimond, M. Brune, and S. Haroche, Rev. Phys.
Mod. \textbf{73}, 565 (2001).

\bibitem{Guo} J.-S. Tang, Y.-L. Li, C.-F. Li, and G.-C. Guo, arXiv:1204.5304
{[}quant-ph{]} (2012).

\bibitem{Werner} R. F. Werner, Phys. Rev. A \textbf{40}, 4277 (1989).

\bibitem{Gleyzes07} S. Gleyzes, S. Kuhr, C. Guerlin, J. Bernu, S.
Deléglise, U. B. Hoff, M. Brune, J.-M. Raimond, and S. Haroche, Nature
\textbf{446,} 297 (2007). 
\end{thebibliography}
\end{document}